\documentclass[12pt]{article}
\usepackage{graphicx}
\usepackage{amsmath}
\usepackage{amsfonts}
\usepackage{amssymb}

\begin{document}
\begin{flushright}
CAMS/03-02\\
\end{flushright}
\begin{center}
\renewcommand{\thefootnote}{\fnsymbol{footnote}}
{\Large \textbf{Noncommutative Gravity}} \vskip20mm {\large \textbf{{Ali H.
Chamseddine \footnote{{\large \textbf{Presented at TH-2002, Paris, France,
July 2002.}}}}}} \renewcommand{\thefootnote}{\arabic{footnote}} \vskip2cm
\textit{Center for Advanced Mathematical Sciences (CAMS) and\newline Physics
Department, American University of Beirut, Lebanon.\newline }
\end{center}

\vskip2cm

\begin{center}
\textbf{Abstract}
\end{center}

\vskip.3cm Various approaches by the author and collaborators to define
gravitational fluctuations associated with a noncommutative space are
reviewed. \vskip2cm

Geometry of a noncommutative space is defined by the data $\left(
\mathcal{A},H,D\right)  $ where $\mathcal{A}$ is a noncommutative involutive
algebra, $H$ is a separable Hilbert space and $D$ a self-adjoint operator on
$H$ referred to as Dirac operator \cite{AC1}. Geometry on Riemannian manifolds
could be recovered by specializing to the data
\[
\mathcal{A}=C^{\infty}\left(  M\right)  ,\quad H=L^{2}(S),\quad D=\gamma^{\mu
}\left(  \partial_{\mu}+\frac{1}{4}\omega_{\mu}^{\;ab}\gamma_{ab}\right)  ,
\]
where $\omega_{\mu}^{\;ab}$ is the spin-connection on a manifold $M.$ To
deserve the name geometry the operator $D$ should satisfy certain conditions
\cite{AC2}.

At present there are only few noncommutative spaces which are well understood
such as the noncommutative space of the standard model, the noncommutative
torus, deformed plane $R_{\theta}^{n}$ and the noncommutative spheres $S^{2},$
$S^{3},$ $S^{4}.$ It is relatively easy to develop gauge theories on
noncommutative spaces. To do this we first define the one-form
\[
\rho=\sum_{i}a^{i}db^{i},\quad a^{i},b^{i}\in\mathcal{A},
\]
then define an involutive representation $\pi$ of $\mathcal{A}$ on $H$ such
that
\[
\pi\left(  \sum\limits_{i}a_{0}da_{1}^{i}\cdots da_{n}^{i}\right)
=\sum\limits_{i}a_{0}^{i}\left[  D,a_{1}^{i}\right]  \cdots\left[  D,a_{n}%
^{i}\right]  .
\]
The curvature is defined by $\theta=d\rho+\rho^{2}$ and integration by
\[
\int\alpha=Tr_{\omega}\left(  \pi\left(  \alpha\right)  \left|  D\right|
^{-d}\right)  ,
\]
where $Tr_{\omega}$ is the Dixmier trace and $d$ is defined by the condition
$tr_{H}\left(  D^{2}+1\right)  ^{-p}<\infty,$ $\forall$ $p>\frac{d}{2}.$

For gauge theories we consider $\alpha=\theta^{2}.$ On a commutative space one
gets, in $d=4$, the action
\[
\int d^{4}x\,Tr\left(  F_{\mu\nu}F^{\mu\nu}\right)  .
\]
On the noncommutative torus \cite{AC3}  the triplet is taken to be  $\left(
l\left(  \mathcal{A}\right)  ,\,l\left(  H\right)  ,\,D\right)  ,$ where $l$
is the left-twisting operator satisfying  $l(a.b)=l(a)\ast l(b)$ \cite{Landi}.
\ The star product is defined by
\[
f\ast g=e^{\frac{i}{2}\theta^{\mu\nu}\frac{\partial}{\partial\xi^{\mu}}%
\frac{\partial}{\partial\eta^{\nu}}}f\left(  x+\xi\right)  g\left(
x+\eta\right)  |_{\xi=\eta=0}.
\]
Then one gets \cite{SW}
\[
\int d^{4}x\,Tr\left(  F_{\mu\nu}\ast F^{\mu\nu}\right)  ,
\]
where $F_{\mu\nu}=\partial_{\mu}A_{\nu}-\partial_{\nu}A_{\mu}+A_{\mu}\ast
A_{\nu}-A_{\nu}\ast A_{\mu}$.

The operator $D$ includes the metric properties  on the space. One can extract
dynamics of fluctuations of the metric by one of two possibilities. The first
is by using the spectral action principle which states that the physical
action depends on the spectrum of $D$ \cite{ACAC}. Good tests of this
principle can be made by considering the standard model of particle physics
and loop space of superstrings. As an example consider the noncommutative
space of the spectral model defined by%
\[
\mathcal{A}=\mathcal{A}_{1}\otimes\mathcal{A}_{2},\quad\,H=H_{1}\otimes
H_{2},\quad D=D_{1}\otimes1+\gamma_{5}\otimes D_{2},
\]
where $\left(  \mathcal{A}_{1},H_{1},D_{1}\right)  $ is the triple associated
with the Riemannian manifold $M$, $\left(  \mathcal{A}_{2},H_{2},D_{2}\right)
$ is the triple associated with the discrete space
\[
\mathcal{A}_{2}=\mathbb{C\oplus M}_{2}\left(  \mathbb{C}\right)
\oplus\mathbb{M}_{3}\left(  \mathbb{C}\right)  ,
\]
$H_{2}$ is enumerated by quarks and leptons and $D_{2}$ contains information
about the Yukawa couplings. The operator $D$ satisfies the property that if
$\psi\in H$ then the fermionic action will be given by
\[
I_{f}=\left(  \psi,\,D\psi\right)
\]
which includes all fermionic interaction terms. The bosonic action is then
given by
\[
I_{b}=Tr\left(  F\left(  \frac{D^{2}}{\Lambda^{2}}\right)  \right)  ,
\]
where $\Lambda$ is a cut-off scale. At low energies, the arbitrariness in the
choice of the function $F$ would only reflect itself in having few measurable
parameters. We can use heat kernel methods to evaluate the above trace. For
example in $d=4$ we first write $F(P)=\sum\limits_{s}f_{s}P^{s}$ and use the
identity $Tr(P^{-s})=\frac{1}{\Gamma\left(  s\right)  }\int\limits_{0}%
^{\infty}dtt^{s-1}Tre^{-tP}$ and the expansion $Tre^{-tP}\simeq\sum
\limits_{n\geq0}t^{\frac{n}{2}-2}\int\limits_{M}a_{n}\left(  x,P\right)
dv\left(  x\right)  $ to show that%
\[
I_{b}=f_{-2}a_{0}+f_{0}a_{2}+f_{2}a_{4}+\cdots,
\]
where the $a_{n}$ are the Seeley-deWit coefficients \cite{Gilkey} and
\[
f_{-2}=\int\limits_{0}^{\infty}\,u\,F(u)du,\quad f_{0}=\int\limits_{0}%
^{\infty}F(u)du,\quad f_{2}=F^{\prime}(0),\cdots.
\]
From the structure of the quarks and leptons denoted by $Q=(u_{L},d_{L}%
,d_{R},u_{R})$ and $L=(\nu_{L},e_{L},e_{R})$ one can determine the discrete
triple $\left(  \mathcal{A}_{2},H_{2},D_{2}\right)  $ that will give rise to
the Higgs field. For the leptonic sector one obtains
\[
D_{L}=\left(
\begin{array}
[c]{cc}%
\gamma^{\mu}\left(  D_{\mu}-\frac{i}{2}g_{2}A_{\mu}^{\alpha}\sigma^{\alpha
}+\frac{i}{2}g_{1}B_{\mu}\right)   & \gamma_{5}k^{e}H\\
\gamma_{5}k^{\ast e}H^{\dagger} & \gamma^{\mu}\left(  D_{\mu}+ig_{1}B_{\mu
}\right)
\end{array}
\right)  ,
\]
and a similar expression for the Dirac operator of the quarks sector. The
bosonic action is then given by
\begin{align*}
I_{b} &  =\int d^{4}x\sqrt{g}\left(  a\Lambda^{4}+b\Lambda^{2}\left(  \frac
{5}{4}R-2y^{2}H^{\dagger}H\right)  \right.  \\
&  \hspace{0.5in}+c\left(  -18C_{\mu\nu\rho\sigma}C^{\mu\nu\rho\sigma}%
+3y^{2}\left(  D_{\mu}H^{\dagger}D^{\mu}H-\frac{1}{6}RH^{\dagger}H\right)
+\frac{5}{3}g_{1}^{2}B_{\mu\nu}B^{\mu\nu}\right.  \\
&  \hspace{0.6in}\quad\quad\left.  \left.  +g_{2}^{2}F_{\mu\nu}^{\alpha
}F^{\alpha\mu\nu}+g_{3}^{2}G_{\mu\nu}^{i}G^{i\mu\nu}+3z^{2}\left(  H^{\dagger
}H\right)  ^{2}\right)  \right)  +O(\frac{1}{\Lambda^{2}}),
\end{align*}
where $a,$ $b$, $c$ are linearly related to $f_{-2},$ $f_{0}$, $f_{2}$
respectively, and $y$ and $z$ are functions of the Yukawa couplings. By
normalizing the kinetic energies of the gauge fields, one obtains a relation
between the gauge coupling constants of $SU(3)$, $SU(2)$ and $U(1)$ which is
the same as that of $SU(5),$ mainly that $g_{3}^{2}=g_{2}^{2}=\frac{5}{3}%
g_{1}^{2}.$ After rescaling the Higgs field one gets a relation for the Higgs
coupling $\lambda=\frac{4z^{2}}{3y^{4}}g_{3}^{2}.$ Using the fact that the
dominant Yukawa coupling is that of the top quark this relation simplifies to
$\lambda\left(  \Lambda\right)  =\frac{16\pi}{3}\alpha_{3}\left(
\Lambda\right)  .$ Combining this relation with the renormalization group
equations one obtains the bound on the Higgs mass $160$ Gev $<$ m$_{H}<$
$\ 200$ Gev. The unification of the couplings also implies that $\Lambda
\simeq10^{15}$ Gev. The spectral action thus unifies gravity with gauge and
Higgs interactions.

The second possibility is to  study the gravitational field of a
noncommutative space from the structure of the spectral triple by defining the
analogue of Riemannian geometry to be called noncommutative Riemannian
geometry \cite{CFF}. For this one must define connections, curvature, torsion,
etc. For example if $E^{A}$ are basis and $\nabla$ a connection then we have
\begin{align*}
\nabla E^{A} &  =-\Omega_{B}^{A}\otimes E^{B},\quad T(\nabla)E^{A}=T^{A},\\
R(\nabla)E^{A} &  =-\nabla^{2}E^{A}=R_{B}^{A}\otimes E^{B},
\end{align*}
which in component form gives
\[
T^{A}=dE^{A}+\Omega_{B}^{A}E^{B},\quad R_{B}^{A}=d\Omega_{B}^{A}+\Omega
_{C}^{A}\Omega_{B}^{C}.
\]
Applying these definitions to a product of a discrete two point space times a
Riemannian manifold $M$ one finds that the basis is given by $E^{a}=\left(
\begin{array}
[c]{cc}%
\gamma^{a} & 0\\
0 & \gamma^{a}%
\end{array}
\right)  $ and $E^{5}=\left(
\begin{array}
[c]{cc}%
0 & \gamma^{5}\\
-\gamma^{5} & 0
\end{array}
\right)  $ and the Dirac operator by
\[
D=\left(
\begin{array}
[c]{cc}%
\gamma^{a}e_{a}^{\mu}D_{\mu} & \gamma_{5}\phi\\
\gamma_{5}\phi & \gamma^{a}e_{a}^{\mu}D_{\mu}%
\end{array}
\right)  .
\]
The noncommutative Einstein-Hilbert action is
\begin{align*}
I &  =\left\langle E_{A},R_{B}^{A}E^{B}\right\rangle \\
&  =2\int\limits_{M}d^{4}x\sqrt{g}\left(  R-2\partial_{\mu}\sigma\partial
_{\nu}\sigma g^{\mu\nu}\right)  ,
\end{align*}
where $\phi=e^{-\sigma}.$ To this we can add a cosmological constant
$\Lambda\left\langle 1\right\rangle =\Lambda\int d^{4}x\sqrt{g}$ and allow
matrix algebras for the discrete space. This would give rise to gauge fields
with curvature $\theta=d\rho+\rho^{2}$, where $\pi\left(  \rho\right)
=\left(
\begin{array}
[c]{cc}%
\gamma^{\mu}A_{\mu} & \gamma_{5}\phi H\\
\gamma_{5}\phi H^{\dagger} & \gamma^{\mu}A_{\mu}%
\end{array}
\right)  .$ The contribution of the gauge fields to the action is then
\cite{lizzi}%
\begin{align*}
I &  =\int\limits_{M}d^{4}x\sqrt{g}\left(  -\frac{1}{4}Tr\left(  F_{\mu\nu
}F^{\mu\nu}\right)  -\frac{\lambda}{4!}\left(  H^{\dagger}H-m^{2}e^{-2\sigma
}\right)  ^{2}\right.  \\
&  \hspace{0.7in}+D_{\mu}H^{\dagger}D^{\mu}H+\left(  \frac{1}{2}%
+k^{2}H^{\dagger}H\right)  \partial_{\mu}\sigma\partial_{\nu}\sigma g^{\mu\nu
}\\
&  \hspace{0.7in}\left.  +R-2\Lambda+\partial_{\mu}\sigma\partial_{\nu}\left(
H^{\dagger}H\right)  \right)  ,
\end{align*}
which is the same at the Randall-Sundrum model \cite{RS} of $4+1$ dimensional
space with four-dimensional brane boundaries.

Allowing the Dirac operator $D$ to fluctuate on noncommutative spaces based on
deformed spheres or deformed $R^{n}$ poses a challenge. The operator $D$ is
not arbitrary, and the interesting problem to solve is to find whether there
are gravitational fluctuations. \ The presence of a constant background
B-field for D-branes leads to noncommutativity of space-time coordinates which
could be realized by deforming the algebra of functions on the world volume.
There are indications that the gravitational action on noncommutative branes
in presence of constant background B-field is non covariant \cite{Ardalan}.

Deformed gravity could be constructed by using the spectral action if one
knows the form of the deformed Dirac operator $\widetilde{D}$. At present this
is not known, and all one can do is to probe for possibilities. If one assumes
a constant background B-field then the commutator of space-time coordinates
gives%
\[
\left[  x^{\mu},x^{\nu}\right]  =i\theta^{\mu\nu}%
\]
where $\left\langle B_{\mu\nu}\right\rangle =\theta_{\mu\nu}$ and $\theta
^{\mu\rho}\theta_{\rho\nu}=\delta_{\nu}^{\mu}.$ In this case the vielbein
would be deformed to the form%
\[
e_{\mu}^{a}\left(  x,\theta\right)  =e_{\mu}^{a}\left(  x\right)
+i\theta^{\nu\rho}e_{\mu\nu\rho}^{a}\left(  x\right)  +\cdots
\]
the complex part is dependent on $\theta.$ If a metric is defined by
$g_{\mu\nu}=e_{\mu}^{a}\ast e_{\nu a}$ then the metric will be complex. We can
write $g_{\mu\nu}=G_{\mu\nu}+iB_{\mu\nu}$ and then impose the hermiticity of
$g_{\mu\nu}$ which implies that $G_{\mu\nu}$ is a symmetric tensor and
$B_{\mu\nu}$ is antisymmetric tensor. This leads to complex gravity. At the
linearized level, the field $B_{\mu\nu}$ has the correct kinetic terms, but
the ghost modes present in \ this field propagate at the non-linear level. The
reason for the inconsistency is that there is no symmetry similar to
diffeomorphism invariance associated with the field $B_{\mu\nu}$ \cite{ali1}.

An alternative approach is to develop a gauge theory for the field $e_{\mu
}^{a}$ and formulate gravity as a noncommutative gauge theory without using a
metric. This can be done without much complications in four-dimensions. The
procedure is generalizable to higher dimensions, but the analysis will be more
complicated. The idea is to start from the gauge group $U(2,2)$ in
four-dimensions. The gauge field is expanded in the form \cite{ali2}%
\[
A_{\mu}=ia_{\mu}+b_{\mu}\Gamma_{5}+e_{\mu}^{a}\Gamma_{a}+f_{\mu}^{a}\Gamma
_{a}\Gamma_{5}+\frac{1}{4}\omega_{\mu}^{ab}\Gamma_{ab}.
\]
The gauge field strength is given by $F=dA+A^{2}$ where we have defined
$A=A_{\mu}dx^{\mu}$ and $F=\frac{1}{2}F_{\mu\nu}dx^{\mu}\wedge dx^{\nu}.$ To
make the system dynamical without using a metric a constraint is imposed. It
is given by%
\[
F_{\mu\nu}^{a}+F_{\mu\nu}^{a5}=0.
\]
This breaks the symmetry to $SL(2\mathbb{C)}$, which is the relevant symmetry
for gravity. The constraint is solved by
\[
f_{\mu}^{a}=\alpha e_{\mu}^{a},\quad\omega_{\mu}^{ab}=\omega_{\mu}^{ab}\left(
e_{\mu}^{a}+f_{\mu}^{a}\right)  ,\quad b_{\mu}=0.
\]
An action which is invariant under the surviving $SL(2\mathbb{C)}$ is given by%
\begin{align*}
I &  =i\int Tr\left(  \Gamma_{5}F\wedge F\right)  \\
&  =\frac{i}{4}\int d^{4}x\,\epsilon^{\mu\nu\rho\sigma}\epsilon_{abcd}\left(
R_{\mu\nu}^{ab}+8\left(  1-\alpha^{2}\right)  e_{\mu}^{a}e_{\nu}^{b}\right)
\left(  R_{\rho\sigma}^{cd}+8\left(  1-\alpha^{2}\right)  e_{\rho}%
^{c}e_{\sigma}^{d}\right)  .
\end{align*}
This action consists of a Gauss-Bonnet topological term, an Einstein term and
a cosmological term, provided that $\alpha\neq1.$ The advantage of this
formulation is that it immediately generalizes to a noncommutative action as
the metric is not used in the construction, but is generated as a function of
$e_{\mu}^{a}.$ Let $\widetilde{A}$ $=\widetilde{A}_{\mu}^{I}T_{I}dx^{\mu}$ be
the gauge field over the noncommutative space where $T_{I}$ are the group
generators. The curvature is then given by $\widetilde{F}=d\widetilde
{A}+\widetilde{A}\ast\widetilde{A}.$ Notice that we can write
\[
\widetilde{A}\ast\widetilde{A}=\frac{1}{2}\left(  \widetilde{A}_{\mu}^{I}%
\ast_{s}\widetilde{A}_{\nu}^{J}\left[  T_{I},T_{J}\right]  +\widetilde{A}%
_{\mu}^{I}\ast_{a}\widetilde{A}_{\nu}^{J}\left\{  T_{I},T_{J}\right\}
\right)  dx^{\mu}\wedge dx^{\nu},
\]
where the symmetric and antisymmetric star products are defined to be,
respectively, even and odd functions of $\theta.$ Notice that this is
consistent with imposing the constraint%
\[
\widetilde{F}_{\mu\nu}^{a}+\widetilde{F}_{\mu\nu}^{a5}=0,
\]
because these correspond to generators with an odd number of gamma matrices
and preserve the subgroup $SL(2\mathbb{C)}$. The noncommutative action is then
given by%
\begin{align*}
I &  =i\int Tr\left(  \Gamma_{5}F\ast F\right)  \\
&  =i\int d^{4}x\,\epsilon^{\mu\nu\rho\sigma}\left(  2\widetilde{F}_{\mu\nu
}^{1}\ast_{s}\widetilde{F}_{\rho\sigma}^{5}+\epsilon_{abcd}\widetilde{F}%
_{\mu\nu}^{ab}\ast_{s}\widetilde{F}_{\rho\sigma}^{cd}\right)
\end{align*}
The constraints could be solved by using the Seiberg-Witten map \cite{SW}
which enables us to express all the deformed fields in terms of the undeformed
ones. The final result is \cite{ali2}
\[
I=i\int d^{4}x\epsilon^{\mu\nu\lambda\sigma}\left(  \epsilon_{abcd}F_{\mu\nu
}^{ab}F_{\lambda\sigma}^{cd}+\theta^{\kappa\rho}\left(  2e_{\mu}^{a+}e_{\nu
a}^{-}F_{\lambda\sigma\kappa\rho}^{1}+\epsilon_{abcd}F_{\mu\nu}^{ab}%
F_{\lambda\sigma\kappa\rho}^{cd}\right)  \right)  +O(\theta^{2})
\]
where $F_{\lambda\sigma\kappa\rho}^{cd}$ is the deformation to first order in
$\theta$ of $F_{\lambda\sigma}^{cd}.$

This deformed action gives deviations from the Einstein-action which are
evaluated explicitly. In this deformation there are no additional propagating
degrees of freedom, all new terms being functions of the vierbein $e_{\mu}%
^{a}$ and its derivatives. This suggests that there should exist a formulation
of noncommutative gravity on deformed $R_{\theta}^{n}$ obtained by allowing
fluctuations to the Dirac operator in the spectral triple of this space. The
nature of such extension is presently under investigation.

To conclude, it is clear from the above discussion that gauge theories on
noncommutative spaces are straightforward, but to define the gravitational
action on such spaces is more difficult. At present there are only partial
answers, and more intensive research in this direction is needed.

\end{document}